\documentstyle[epsfig]{aprim10}
\input{epsf}

\newif\ifAMStwofonts

\ifoldfss
  \ifCUPmtlplainloaded \else
    \NewTextAlphabet{textbfit} {cmbxti10} {}
    \NewTextAlphabet{textbfss} {cmssbx10} {}
    \NewMathAlphabet{mathbfit} {cmbxti10} {} 
    \NewMathAlphabet{mathbfss} {cmssbx10} {} 
  \fi
  \ifAMStwofonts
    \ifCUPmtlplainloaded \else
      \NewSymbolFont{upmath} {eurm10}
      \NewSymbolFont{AMSa} {msam10}
      \NewMathSymbol{\upi}     {0}{upmath}{19}
      \NewMathSymbol{\umu}     {0}{upmath}{16}
      \NewMathSymbol{\upartial}{0}{upmath}{40}
      \NewMathSymbol{\leqslant}{3}{AMSa}{36}
      \NewMathSymbol{\geqslant}{3}{AMSa}{3E}

    \fi
  \fi
\fi 

\ifnfssone
  \newmathalphabet{\mathit}
  \addtoversion{normal}{\mathit}{cmr}{m}{it}
  \addtoversion{bold}{\mathit}{cmr}{bx}{it}
  \newmathalphabet{\mathbfit} 
  \addtoversion{normal}{\mathbfit}{cmr}{bx}{it}
  \addtoversion{bold}{\mathbfit}{cmr}{bx}{it}
  \newmathalphabet{\mathbfss} 
  \addtoversion{normal}{\mathbfss}{cmss}{bx}{n}
  \addtoversion{bold}{\mathbfss}{cmss}{bx}{n}
  \ifAMStwofonts
    \ifCUPmtlplainloaded \else
      \UseAMStwoboldmath
      \makeatletter
      \new@mathgroup\upmath@group
      \define@mathgroup\mv@normal\upmath@group{eur}{m}{n}
      \define@mathgroup\mv@bold\upmath@group{eur}{b}{n}
      \edef\UPM{\hexnumber\upmath@group}
      \new@mathgroup\amsa@group
      \define@mathgroup\mv@normal\amsa@group{msa}{m}{n}
      \define@mathgroup\mv@bold\amsa@group{msa}{m}{n}
      \edef\AMSa{\hexnumber\amsa@group}
      \makeatother
      \mathchardef\upi="0\UPM19
      \mathchardef\umu="0\UPM16
      \mathchardef\upartial="0\UPM40
      \mathchardef\leqslant="3\AMSa36
      \mathchardef\geqslant="3\AMSa3E
    \fi
  \fi
\fi 

\ifnfsstwo
  \DeclareMathAlphabet{\mathbfit}{OT1}{cmr}{bx}{it}
  \SetMathAlphabet\mathbfit{bold}{OT1}{cmr}{bx}{it}
  \DeclareMathAlphabet{\mathbfss}{OT1}{cmss}{bx}{n}
  \SetMathAlphabet\mathbfss{bold}{OT1}{cmss}{bx}{n}
  \ifAMStwofonts
    \ifCUPmtlplainloaded \else
      \DeclareSymbolFont{UPM}{U}{eur}{m}{n}
      \SetSymbolFont{UPM}{bold}{U}{eur}{b}{n}
      \DeclareSymbolFont{AMSa}{U}{msa}{m}{n}
      \DeclareMathSymbol{\upi}{0}{UPM}{"19}
      \DeclareMathSymbol{\umu}{0}{UPM}{"16}
      \DeclareMathSymbol{\upartial}{0}{UPM}{"40}
      \DeclareMathSymbol{\leqslant}{3}{AMSa}{"36}
      \DeclareMathSymbol{\geqslant}{3}{AMSa}{"3E}
    \fi
  \fi
\fi 

\ifCUPmtlplainloaded \else
  \ifAMStwofonts \else 
    \def\upi{\pi}
    \def\umu{\mu}
    \def\upartial{\partial}
  \fi
\fi

\title[Chromospheric activity]{Chromospheric activity on the RS Canum Venaticorum stars}

\author[Zhang L. Y., \& Gu S. H.,]
       {L. Y. Zhang$^{1,2}$ and S. H. Gu$^2$\\
        $^1$College of Science/Department of Physics, Guizhou University,
Guiyang 550025, PR China\\
        $^2$National Astronomical Observatories/Yunnan
   Observatory,
       Chinese Academy of Sciences, Kunming 650011, PR China}
\date{}

\pagerange{\pageref{firstpage}--\pageref{lastpage}}
\pubyear{2008}

\begin{document}

\maketitle

\label{firstpage}

\begin{abstract}
Firstly, we review the stellar chromospheric activity in the optical
wavelength. Secondly, we introduce our ongoing project of
multi-wavelength high--resolution optical observations aimed at
studying the chromospheric activity of different RS CVn stars.
Finally, we give our future perspectives.
\end{abstract}
\section{Introduction}
This is a brief review of stellar chromospheric activity. In this
section, we will discuss the definition of chromosphere,
chromospheric activity, chromospheric
activity indicators and diagnostic technique.\\
\textbf{1.1 What is a chromosphere}\\
\indent In classical, the chromosphere is an intermediate region in
the atmosphere of a star, lying above the photosphere and below the
corona. At present, Hall (2008) gave us a working definition of the
chromosphere. It is the region of a stellar atmosphere where we
observe emission in excess of that expected in radiative equilibrium
and where cooling occurs mainly due to radiation in strong resonance
lines (rather than in the continuum, mostly the case in the photosphere) of abundant species such as $\mbox{Mg~{\sc ii}}$ and $\mbox{Ca~{\sc ii}}$.\\
\textbf{1.2 Chromospheric activity}\\
\indent For late-type stars with thick convective zones and rapid
rotation, they exhibit chromospheric activity phenomena such as
plage and flare, which are tightly linked to changes of the stellar
magnetic field. Chromospheric plage produces emission in the cores
of the $\mbox{Ca~{\sc ii}}$ H\&K lines (Fig.1). For the observations
of HR 1099 obtained by Garc\'{i}a-Alvarez et al (2003), optical
flare was detected (see Fig. 2). The equivalent widths (EWs) of
H$_{\alpha}$ emission increase by almost a factor of 4. \\
\indent By analyzing the chromospheric activity variation with
orbital phase, astronomers have found some stars show rotational
modulation phenomena. Fig.3 shows a example of a clear rotational
modulation of the H$_{\alpha}$ emission of LQ Hya (Frasca et al.
2008). They applied a simple geometric plage model to explain the
rotational modulated chromospheric emission.
\begin{figure} \centering
\includegraphics[width=4.0cm,height=2.5cm]{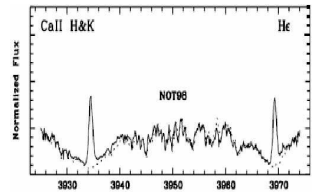}
 \caption{The $\mbox{Ca~{\sc ii}}$ H\&K profiles of $\sigma$ Gem (Montes et al. 2000).}
\end{figure}
\begin{figure}
\centering
\includegraphics[width=4.5cm,height=3.5cm]{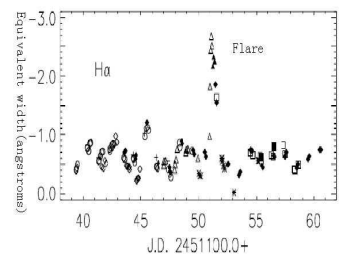}
 \caption{The H$_{\alpha}$ equivalent width as a function of Julian date for HR 1099 (Garc\'{i}a-Alvarez et al. 2003).}
\end{figure}
The most prominent monitoring programme of solar-type chromospheric
activity is called the HK project (Wilson 1963, 1978). Long term
monitoring of chromospheric activity has revealed that many cool
stars
show different activity cycles (see Fig.4) (Berdyugina 2005). \\
\textbf{1.3 Chromospheric activity indicators} \\
\indent Chromospheric activity produce fill-in or emission in some
strong photospheric lines. Usually we use these lines as
chromospheric
activity indicators. These indicators are summarized as follows: \\
\textbf{The $\mbox{Na~{\sc i}}$\ D$_{1}$, D$_{2}$ lines and $\mbox{Mg~{\sc i}}$ b triplet lines:}\\
\indent The $\mbox{Na~{\sc i}}$\ D$_{1}$, D$_{2}$ and $\mbox{Mg~{\sc
ii}}$ b triplet lines are formed in the upper photosphere and lower
chromosphere. Both $\mbox{Na~{\sc i}}$ and $\mbox{Mg~{\sc i}}$ lines
are detected during flares as emission reversal or as filled-in
absorption (Andretta et al. 1997;
Montes et al. 1997; Montes et al. 2004).\\
\textbf{The $\mbox{Ca~{\sc ii}}$ infrared triplet (IRT) lines:}\\
 \indent The $\mbox{Ca~{\sc
ii}}$ IRT lines are important chromospheric activity indicators for
the Sun and late-type stars (Gunn \& Doyle 1997; Montes et al. 1997;
Montes et al. 2000). They are formed in the lower chromosphere,
making them sensitive probes of the temperature minimum region
(Montes et al. 1997). The ratio of excess emission,
$EW_{8542}/EW_{8498}$, is an indicator of the chromospheric
structure (plages, prominences). In solar plages, the values of
$EW_{8542}/EW_{8498}$ are in the range 1.5-3 (Chester 1991). These
low values are also found in other chromospherically active binaries
by other authors: L\'{a}zaro \& Ar\'{e}valo (1997); Montes et al.
(2000) and Gu et al. (2002). While in solar prominences, the values
are about 9 (Chester
1991).\\
\begin{figure}
\centering
\includegraphics[width=5cm,height=3cm]{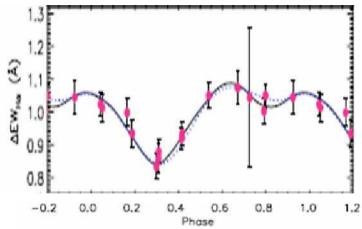}
 \caption{Rotational modulation of the residual H$_{\alpha}$ equivalent width of LQ Hya.
The solid line represents the best fit of 3-plage model, while the
dotted line represents 2-plage model (Frasca et al. 2008) }
\end{figure}
\begin{figure}
\centering
\includegraphics[width=5cm,height=3.5cm]{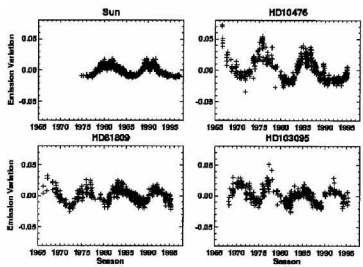}
 \caption{Chromospheric Ca II emission cycles for Solar-like stars, illustrating the regular,
 cyclic variation that is common in such stars. The Ca II emission is plotted in Mount Wilson ¡®S-Index¡¯ units (Radick 2000).}
\end{figure}
\textbf{The H$_{\alpha}$, H$_{\beta}$ and other Balmer lines:}\\
\indent The Balmer lines are very useful indicators of chromospheric
activity and formed in the middle chromosphere (Montes et al. 1997,
etc.). For less active stars, these line profiles are filled-in
absorption. While for much active stars, they are emission above the
continuum. The ratio $EW_{H_{\alpha}}/EW_{H_{\beta}}$ can be used as
a diagnostic indicator for discriminating between plages and
prominences (Hall \& Ramsey 1992; Montes et al., 2004). According to
Buzasi model, the low ratio (1-2) can be achieved both in plages and
prominences viewed against the disk, but the high ratio ($>3$, to a
theoretical maximum of about 15) can only be achieved in extended
regions viewed
off the limb (Hall \& Ramsey 1992).\\
\textbf{The $\mbox{Ca~{\sc ii}}$ H\&K lines:}\\
 \indent The $\mbox{Ca~{\sc ii}}$ H\&K lines have been the traditional diagnostic indicators of
chromospheric activity in cool stars for long time. They are formed
in the middle chromosphere. The emissions in the cores of these
lines are the most widely used optical indicators of chromospheric
activity.\\
\textbf{The $\mbox{He~{\sc i}}$ D$_{3}$ lines:}\\
\indent The $\mbox{He~{\sc i}}$ D$_{3}$ line is formed in the upper
chromosphere. The emission of the line is a probe for detecting
flare-like
events (Zirin 1988).\\
\textbf{1.4 Diagnostic technique}\\
\indent To extract the chromospheric contribution from the spectra
line, the method is the spectral subtraction technique. The
principle of the method is that chromospheric contribution equals to
observed spectra minus the synthesized spectra. The problem of the
technique is to simulate the correct synthesized spectrum
representing the underlying photospheric contribution. There are two
approaches. One is using theoretical spectra based on radiative
transfer solutions of model atmospheres (Fraquelli 1984). The
theoretical line profiles were calculated from the model atmospheres
by using the known effective temperature and surface gravities of
the active system. The problem with theoretical line profiles for
spectral subtraction is the uncertainty and complexity of the
atmospheric conditions. Without detailed information concerning the
dominating effects on the source functions of active lines and the
effects of active regions on these lines, it is impossible to form
adequate theoretical representation of the inactive contribution
(Gunn \& Doyle 1997). The other is using observed spectra of
inactive stars. The synthesized spectrum is constructed from
artificially rotationally broadened, radial-velocity shifted, and
weighted spectra of inactive stars with the same spectral type and
luminosity class as
the components of the active system (Barden 1985). There are some assumptions in this spectral subtraction technique (Barden 1985; Gunn \& Doyle 1997). Usually we use the second method.\\

\section{our ongoing project}
\indent 
We aim to study the chromospheric activity of different RS CVn
stars. By analyzing the simultaneous spectroscopic observations of
several chromospheric activity indicators for the RS CVn binary
systems, and by using the spectral subtraction technique, we have
investigated the detail of the excess emission and studied the
chromospheric activity variation with orbital phase and
different epoches.\\
\indent The main observational method is high-resolution
spectroscopy. The telescope and instrumental configuration is 2.16
meter telescope with echelle spectrograph of Xinglong station, NAOC.
The used wavelength region is from 5600 to 9000 ${\AA}$ and the
resolution is about 37,000.
Up to now, we have analyzed chromospheric activity on the RS CVn
binary SZ Psc (P$_{rot}$=$3^{d}$.97, F8V+K1IV) (Zhang \& Gu 2008).
\begin{figure}
\includegraphics[width=2.75cm,height=2.5cm]{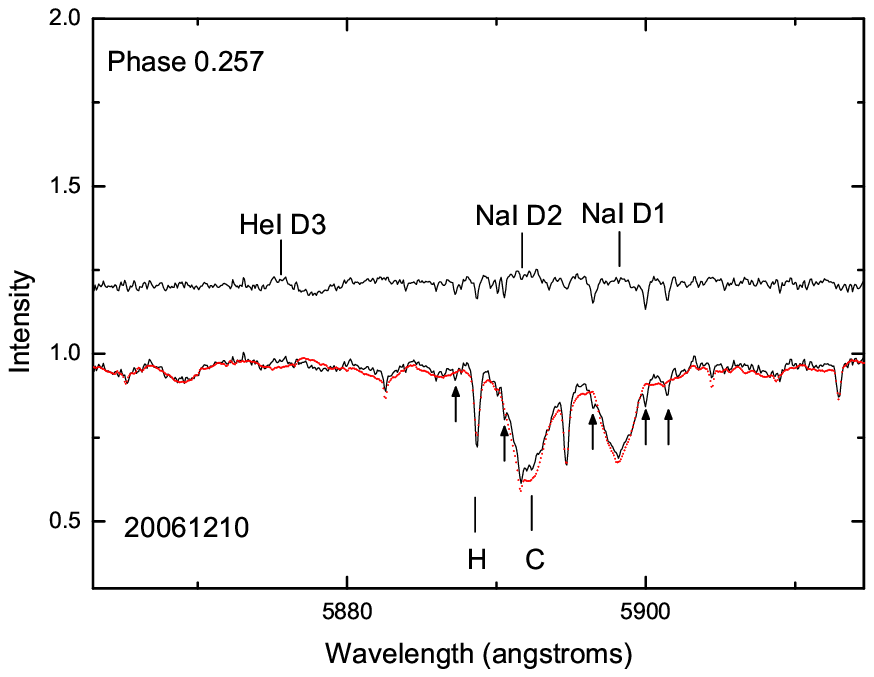}
\includegraphics[width=2.75cm,height=2.5cm]{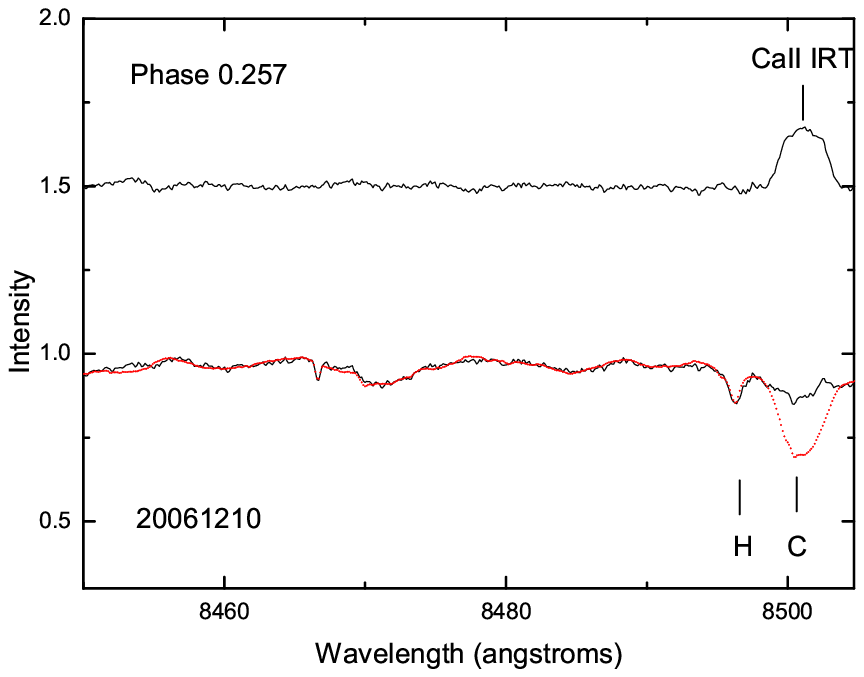}
\includegraphics[width=2.75cm,height=2.5cm]{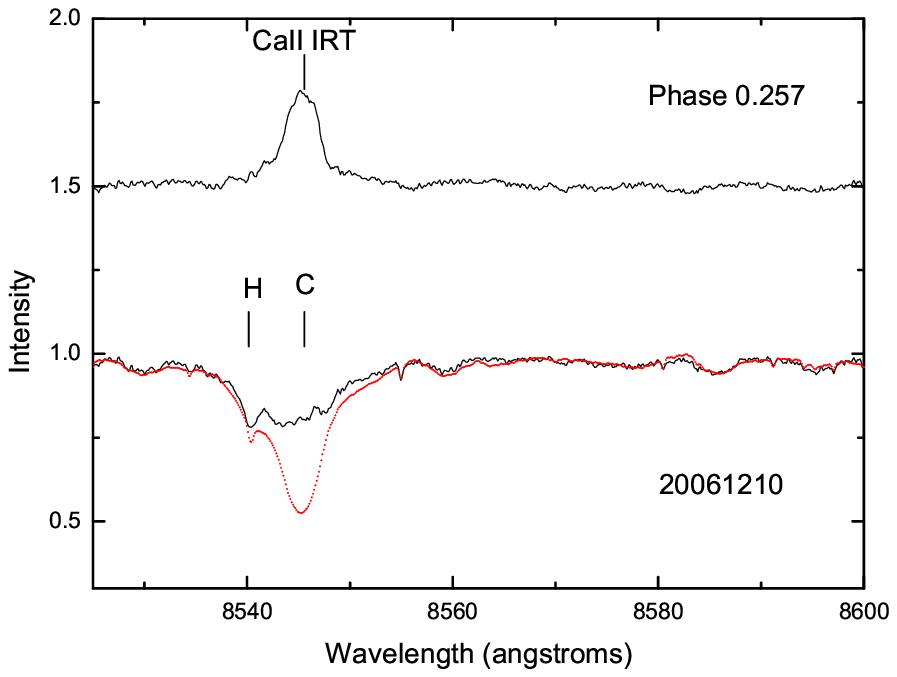}
\includegraphics[width=2.75cm,height=2.5cm]{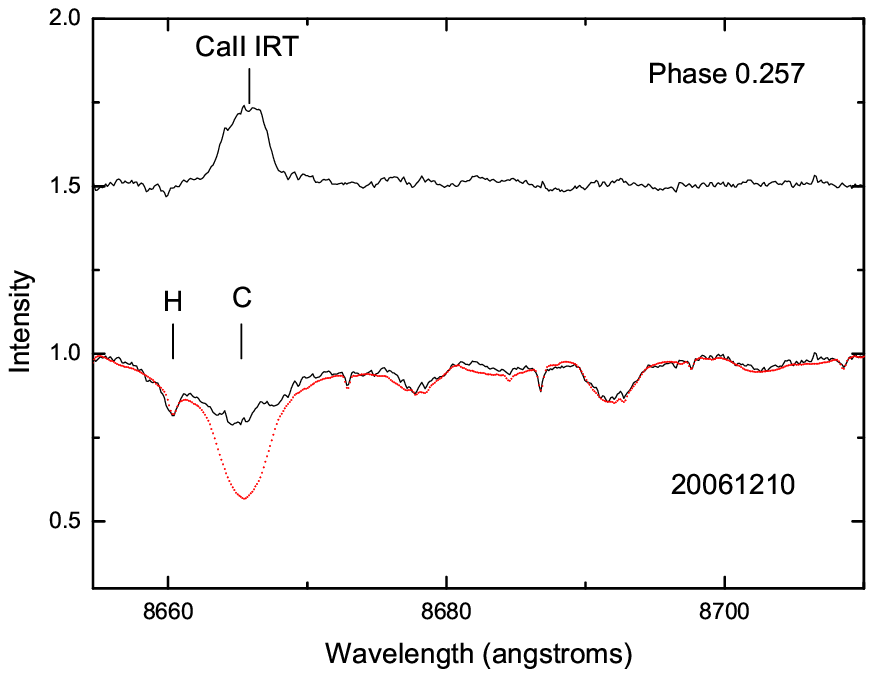}
\includegraphics[width=2.75cm,height=2.5cm]{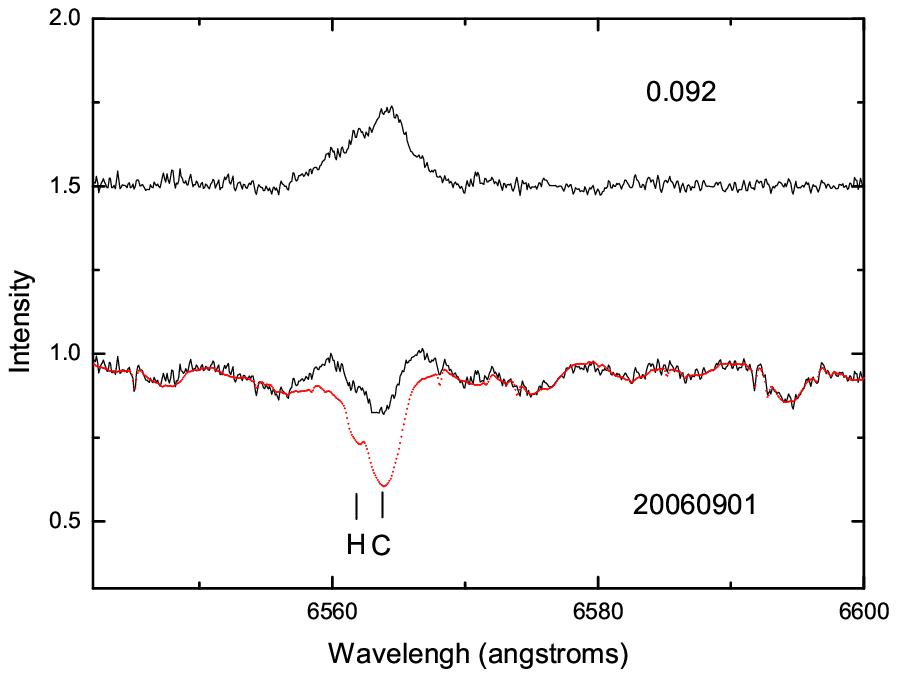}
\includegraphics[width=2.75cm,height=2.5cm]{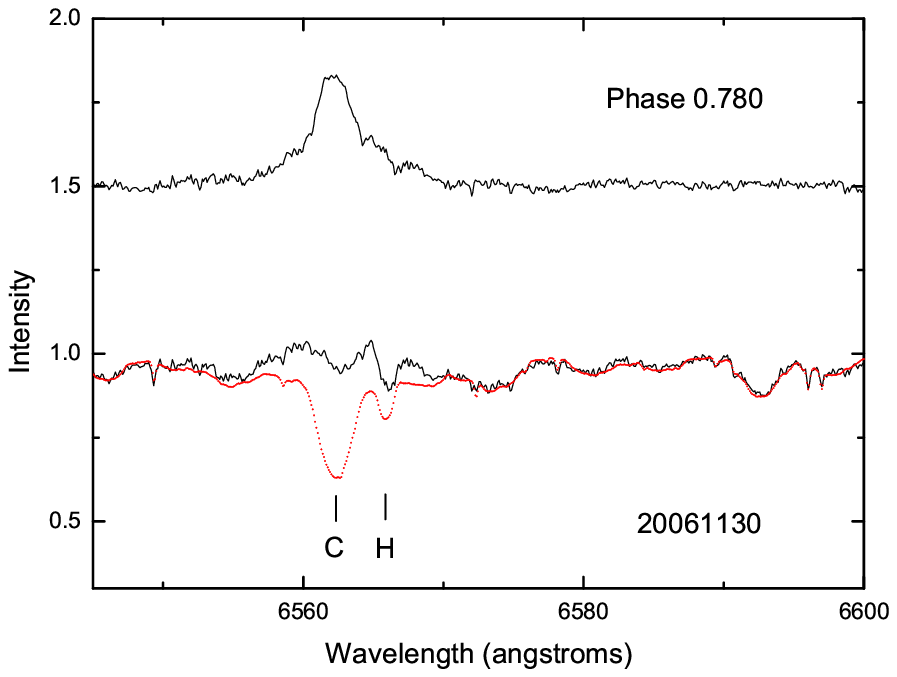}
\caption{Samples of the observed, synthesized, and subtracted
spectra for the $\mbox{He~{\sc i}}$ D$_{3}$, $\mbox{Na~{\sc i}}$
D$_{1}$, D$_{2}$, $\mbox{Ca~{\sc ii}}$ IRT and H$_{\alpha}$ lines.
The dotted lines represent the synthesized spectra and the upper
spectra are the subtracted spectra. Vertical arrows mark the
telluric lines that appeared in the spectral region.}
\end{figure}
Our spectroscopic observations were made in four observing runs:
Sept. 1-6, Oct. 28-29, Nov. 28-30, and Dec. 8-10, 2006. Each
observational run includes the optical chromospheric indicators: the
$\mbox{He~{\sc i}}$ D$_{3}$, $\mbox{Na~{\sc i}}$\, D$_{1}$, D$_{2}$,
H$_{\alpha}$, and $\mbox{Ca~{\sc ii}}$ IRT lines. The method we used
is the spectral subtraction technique and the code is starmod
developed by Barden (Barden 1985). Some examples of the different
chromospheric indicators are displayed in
Fig.5.\\
\indent The application of the spectral subtraction technique
reveals that the $\mbox{Na~{\sc i}}$\, D$_{1}$, D$_{2}$ lines, in
some cases, exhibit obvious excess emission from the cooler
component. For the $\mbox{He~{\sc i}}$ D$_{3}$ line, there is no
obvious absorption or emission. So, during our observing seasons, we
observed no flare-like episodes. For the $\mbox{Ca~{\sc ii}}$ IRT
lines, they show obvious excess emission from the cooler component.
For the H$_{\alpha}$ line, it shows excess emission from the cooler
component and the excess emission profiles exhibit broad wings.
There are a couple of possible explanations for the H$_{\alpha}$
line. First, the broad component could be interpreted as arising
from microflaring. Second, it might be caused by instantaneous mass
transfer from the cooler component to the hotter one(Bopp 1981). In
summary, all these analyzed activity indicators show
that the cooler component is active.\\
\indent We measured the EWs of the excess emissions. To discuss the
rotational modulation of chromospheric activity, we used all these
data because we only have 3 to 4 data points per rotation in
different epochs. Fig.6 shows the EWs with orbital phase. We used
polynomial function to fit the data. For the $\mbox{Ca~{\sc ii}}$
8498 data, there is no significant trend. While for the
$\mbox{Ca~{\sc ii}}$ 8542 and 8662 lines, especially for the
$\mbox{Ca~{\sc ii}}$ 8542 line, it seems that the emission is
stronger near phases 0.25 and 0.75. For the H$_{\alpha}$ line, it
seems that the emissions is
\begin{figure}
\centering
\includegraphics[width=4.17cm,height=3.5cm]{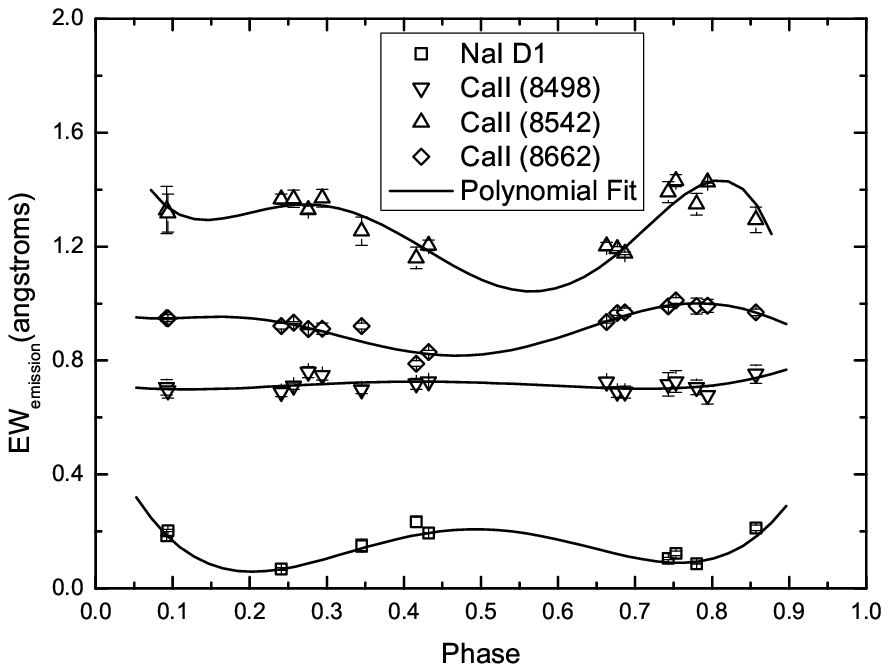}
\includegraphics[width=4.17cm,height=3.5cm]{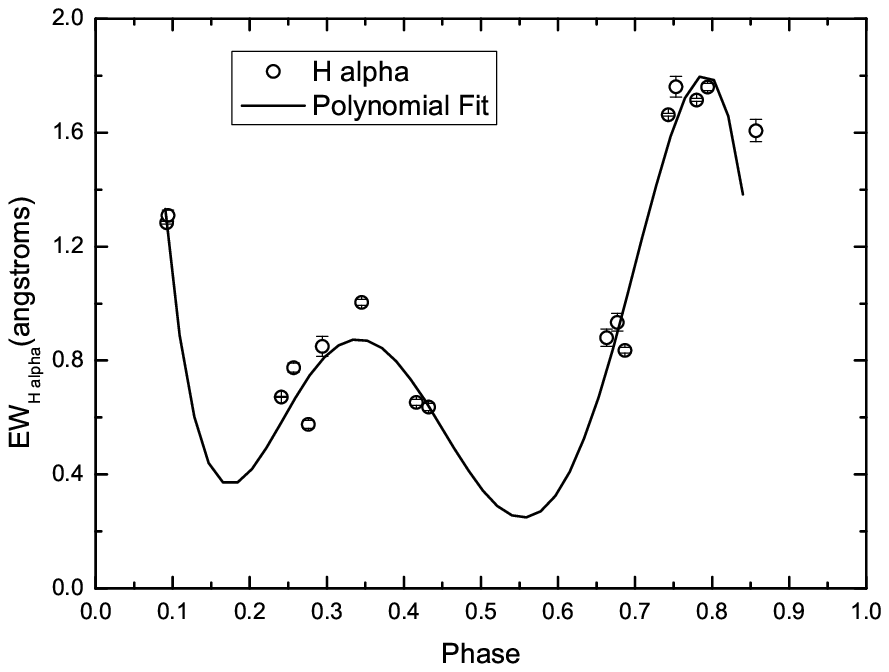}
\caption{The EWs of the excess emissions vs. orbital phase for
$\mbox{Na~{\sc i}}$ D, $\mbox{Ca~{\sc ii}}$ IRT and H$_{\alpha}$
lines. The solid line refers to a polynomial fit to the data.}
\end{figure}
stronger near phase 0.3 and 0.75. Therefore, for the $\mbox{Ca{\sc
ii}}$ 8542 and 8662 and H$_{\alpha}$ lines, the excess emissions
(with the orbital phase) may be correlated basically, especially
around two quadratures. The emissions are stronger around two
quadratures of the system (phases 0.25 and 0.75). However, for the
$\mbox{Na~{\sc i}}$ D$_{1}$ line, the emission is weaker around the
two quadratures, so the $\mbox{Na~{\sc i}}$ D$_{1}$ line may be
anti-correlated with the $\mbox{Ca~{\sc ii}}$
8542, 8662 and the H$_{\alpha}$ lines.\\
\section{Perspective}
We would like to give our future plan, as follows:\\
1. Monitor long-term chromospheric activity of SZ Psc.\\
2. Chromospheric activity studies of other RS CVn stars.\\
3. UV, x-ray and radio studies of selected RS CVn objects.\\
\indent Finally, we will investigate the chromospheric activity
evolution with age and its dependency on stellar parameters such as
stellar rotation, mass and so on. \\
\section*{Acknowledgements}
The authors would like to thank the observing assistants of the 2.16
meter telescope of Xinglong station for their help and support
during our observations. We are very grateful to Dr. Montes for
providing a copy of the STARMOD program. We also would like to thank
Mr. Xiang-song Fang for his valuable suggestions and comments, which
have led to significant improvements in our manuscript. This work is
supported by
the NSFC under grants No. 10373023 and 10773027. 

\label{lastpage}

\clearpage

\end{document}